\documentclass[twoside,english,aps,twocolumn]{revtex4}
\usepackage{graphicx}
\usepackage{amsmath}

\makeatletter

\providecommand{\LyX}{L\kern-.1667em\lower.25em\hbox{Y}\kern-.125emX\@}

\makeatother
\begin{document}

\title{Entanglement Sharing in the Two-Atom Tavis-Cummings Model}

\author{T. E. Tessier}

\email{tessiert@info.phys.unm.edu}

\affiliation{Department of Physics and Astronomy, University of New Mexico, 800
Yale Boulevard, 87131 Albuquerque, USA}

\author{I. H. Deutsch}

\email{ideutsch@info.phys.unm.edu}
\homepage{info.phys.unm.edu/~DeutschGroup}

\affiliation{Department of Physics and Astronomy, University of New Mexico, 800
Yale Boulevard, 87131 Albuquerque, USA}

\author{A. Delgado}

\email{aldo.delgado@udec.cl}

\affiliation{Department of Physics and Astronomy, University of New Mexico, 800
Yale Boulevard, 87131 Albuquerque, USA}

\author{I. Fuentes-Guridi}

\email{ifuentes-guridi@perimeterinstitute.ca}

\affiliation{Perimeter Institute, 35 King Street, North Waterloo, Ontario, Canada
N2J 2W9}

\affiliation{Optics Section, The Blackett Laboratory, Imperial College, London SW7 2BZ, United Kingdom}

\begin{abstract}
Individual members of an ensemble of identical systems coupled to
a common probe can become entangled with one another, even when they
do not interact directly. We investigate how this type of multipartite
entanglement is generated in the context of a system consisting of
two two-level atoms resonantly coupled to a single mode of the electromagnetic
field. The dynamical evolution is studied in terms of the entanglements
in the different bipartite partitions of the system, as quantified
by the I-tangle. We also propose a generalization of the so-called
residual tangle that quantifies the inherent three-body correlations
in our tripartite system. This enables us to completely characterize the phenomenon of entanglement sharing in the case of the two-atom
Tavis-Cummings model, a system of both theoretical and experimental
interest.
\end{abstract}

\date{\today{}}

\maketitle

\section{introduction\label{Sec_Introduction}}

The control of quantum systems through active measurement and feedback
has been developing at a rapid pace. In a typical scenario, a single
atom is monitored indirectly through its coupling to a traveling probe
such as a laser beam. The scattered beam and the system become correlated,
and a subsequent measurement of the probe leads to backaction on
the system. A coherent drive applied to the system can then be made
conditional on the measurement record, leading to a closed-loop control
model \cite{Wiseman94, Doherty00}. Such a protocol has been implemented to control a
single mode electromagnetic field in a cavity \cite{Armen02}, and has been
envisioned for controlling a variety of systems such as the state
of a quantum dot in a solid \cite{Goan01}, the state of an atom coupled
to a cavity mode \cite{Ye99}, and the motion of a micro-mechanical resonator
coupled to a Cooper pair charge box \cite{Armour02}.

A common theme in the examples given above is that measurements are
made on \emph{single} copies of the quantum system of interest. However, in many situations one does not have access to an individually
addressable system. In a gas, for example, preparing and/or addressing
individual atoms is extremely difficult. In situations such as this, it is useful to
think of the entire ensemble as a single many-body system. Indeed,
recent experiments \cite{Kuzmich00, Julsgaard01} and theoretical proposals \cite{Thomsen02} have
explored the control of such ensembles from the point of view of the
Dicke model \cite{Stockton02}, where a collection of \( N \) two-level atoms
is treated as a pseudo-spin with \( J=N/2 \).

Measurement backaction on the pseudo-spin can lead to squeezing of
the quantum fluctuations \cite{Kuzmich00, Julsgaard01, Thomsen02}, which may be enhanced through
active closed-loop control \cite{Wiseman94, Doherty00}. This squeezing can reduce the
quantum fluctuations of an observable as in, for example, the reduction
of {}``projection noise\char`\"{} leading to enhanced precision measurements
in an atomic clock \cite{Wineland94}. Moreover, spin-squeezing is related
to quantum entanglement between the atomic members of the ensemble
\cite{Sorensen01}. This entanglement arises not through direct interaction
between the atoms, but through their coupling to a common {}``quantum
bus\char`\"{} in the form of an applied probe.

Measures of entanglement associated with these spin-squeezed
states have been studied by Stockton, \emph{et. al.,}
\cite{Stockton02} under the assumption that all of the atoms in the
ensemble are symmetrically coupled to the bus. However, completely
quantifying entanglement in the most general cases is extremely
difficult, and as yet, an unsolved problem \cite{Bruss02}. In
this article we consider the simplest possible ensemble consisting
of two two-level atoms. Although at first sight this might appear
trivial, when such a system is coupled to a quantum bus a rich
structure emerges. Again, we consider the simplest realization of
the bus -- a single mode quantized electromagnetic field. The
resulting physical system then corresponds to the two-atom
Tavis-Cummings model (TCM) \cite{Tavis68}. A thorough understanding
of the dynamical evolution of the TCM has obvious implications for
the performance of quantum information processing
\cite{Nielsen00,Alber01,Lo98}, as well as for our understanding of
fundamental quantum mechanics \cite{Nielsen00,Peres95}.  Bipartite
entanglement has been investigated in this system for the one-atom
case, known as the Jaynes-Cummings model, for initial pure states
\cite{Loudon87} and mixed states \cite{Bose01, Scheel02} of the field.

Taken as a whole, the two-atom TCM in an overall pure state constitutes
a \emph{tripartite} quantum system in a Hilbert space with tensor
product structure \( 2\otimes 2\otimes \infty  \). Entanglement in
tripartite systems has been studied by Coffman, \emph{et. al.}, \cite{Coffman00}
for the case of three qubits. They found that such quantum correlations
cannot be arbitrarily distributed amongst the subsystems; the existence
of three-body correlations constrains the distribution of the bipartite
entanglement which remains after tracing over any one of the qubits.
For example, in a GHZ-state, \( |\mathrm{GHZ}\rangle =1/\sqrt{2}(|000\rangle +|111\rangle  \)),
tracing over one qubit results in a maximally mixed state containing
no entanglement between the remaining two qubits. In contrast, for
a W-state, \( |\mathrm{W}\rangle =1/\sqrt{3}(|001\rangle +|010\rangle +|100\rangle  \)),
the average remaining bipartite entanglement is maximal \cite{Dur00}.
Coffman, \emph{et. al.,} analyzed this phenomenon of \emph{entanglement
sharing} \cite{Coffman00}, using an entanglement monotone known as
the tangle, a simple generalization of the more familiar concurrence
\cite{Hill97,Wootters98}. They also introduced a new quantity known as
the residual tangle, in order to quantify the irreducible tripartite
correlations in a three qubit system \cite{Coffman00}.

In this article we extend the analysis of entanglement sharing to
the case of the two-atom TCM. This has implications for the study
of quantum control of ensembles. For example, if we imagine that the
quantum bus is measured, \emph{e.g.}, the field leaks out of the cavity and
is then detected, then the degree of correlation between the field
and one of the atoms determines the degree of backaction on \emph{one}
atom. We can then quantify the degree to which one can perform quantum
control on a single member of an ensemble even when one couples only
\emph{collectively} to the entire ensemble. We will accomplish this
by extending the residual tangle formalism of Coffman, \emph{et. al.},
to our \( 2\otimes 2\otimes \infty  \) system.

The remainder of this article is organized as follows. The important
features of the TCM are reviewed in Sec. \ref{Sec_TCM}, and the applicable
measures of entanglement are introduced in Sec. \ref{Sec_Tangle_Formalism}.
With this formalism in hand, we calculate the tangle for each of the
bipartite partitions of this tripartite system in Sec. \ref{Sec_Bipartite_Tangles}.
We will find an approximate analytic expression for the tangle between
the field and the ensemble in the limit of large average photon number
and in the Markoff approximation which provides further insight into
these results. In Sec. \ref{Sec_Ent_Sharing}, we study the irreducible
tripartite correlations in the system using our proposed generalization
of the residual tangle.  Finally, we summarize our results in section \ref{Sec_Summary}.

\section{The Tavis-Cummings Model\label{Sec_TCM}}

The Tavis-Cummings model (TCM) \cite{Tavis68} (also known as the {}``Dicke
model\char`\" \cite{Mandel95}) describes the simplest fundamental interaction
between a single mode of the quantized electromagnetic field and a
collection of \( N \) atoms under the usual two-level and rotating
wave approximations.  The two-atom \( \left( N=2\right)  \)
TCM is governed by the Hamiltonian

\begin{eqnarray}
H & = & H_{0}+H_{int}\nonumber \\& = & \hbar \omega \left(a^\dagger a+\frac{1}{2}\sigma _{z}^{\left(1\right)}+\frac{1}{2}\sigma _{z}^{\left(2\right)}\right)\nonumber \\
 &  & +\hbar g\left[\left(\sigma _{-}^{\left(1\right)}+\sigma _{-}^{\left(2\right)}\right)a^\dagger +\left(\sigma _{+}^{\left(1\right)}+\sigma _{+}^{\left(2\right)}\right)a\right],\label{Eq TCM Hamiltonian}
\end{eqnarray}

\noindent where \( \sigma _{\pm }^{(i)} \) and
\( \sigma _{z}^{(i)} \) display a local \( SU(2) \) algebra for
the \( i \)-th atom in the two-dimensional subspace spanned by the ground and
excited states \( \{|g\rangle ,|e\rangle \} \), and \( a\left( a^{\dagger }\right)  \)
are bosonic annihilation (creation) operators for the monochromatic
field. The Hilbert space \( {\mathcal{H}} \) of the joint system
is given by the tensor product \( {\mathcal{H}}_{A_{1}}\otimes {\mathcal{H}}_{A_{2}}\otimes {\mathcal{H}}_{F} \)
where \( {\mathcal{H}}_{A_{1}} \) (\( {\mathcal{H}}_{A_{2}} \))
denotes the Hilbert space of atom one (two) and \( {\mathcal{H}}_{F} \)
is the Hilbert space of the electromagnetic field.

The total number of excitations \( K=a^{\dagger }a+\frac{1}{2}(\sigma _{z}^{(1)}+\sigma _{z}^{(2)}+2) \)
is a conserved quantity which allows one to split the Hilbert space
\( {\mathcal{H}} \) into a direct sum of subspaces, \emph{i.e.},
\( {\mathcal{H}}=\sum _{K=0}^{\infty }\oplus \Omega _{K} \), with
each subspace \( \Omega _{K} \) spanned by the eigenvectors \( \{|ee,k-2\rangle ,|eg,k-1\rangle ,|ge,k-1\rangle ,|gg,k\rangle \} \)
of \( K \) with eigenvalue \( k \).  The analytic form for the time
evolution operator within a subspace $\Omega _{K}$ is given in Zeng, \emph{et. al.}, \cite{Zeng01}.

It is assumed throughout that the initial state of the TCM system
is pure. Furthermore, we consider only the effects of the unitary
evolution generated by Eq. (\ref{Eq TCM Hamiltonian}), \emph{i.e.},
we do not include the effects of measurement, nor of mixing due to
environment-induced decoherence \cite{Paz02}, so that our system
remains in an overall pure state at all times. Finally, by assuming
an identical coupling constant \( g \) between each of the atoms
and the field, the Hamiltonian is symmetric under atom-exchange. This
invariance under the permutation group was used by Stockton, \emph{et.
al.,} \cite{Stockton02} to analyze the entanglement properties of
very large ensembles. We will also make use of
this fact in order to reduce the number of different partitioning
schemes that one needs to consider when studying entanglement sharing in the two-atom TCM.

\section{The Tangle Formalism\label{Sec_Tangle_Formalism}}

The tangle between two qubits in an arbitrary state is defined in
terms of the concurrence \cite{Hill97,Wootters98}. For a pure state
$\left|\psi \right\rangle $ of two qubits, the concurrence is given
by $C\left(\psi \right)\equiv \left|\left\langle \psi |\widetilde{\psi }\right\rangle \right|$,
 where $\left|\widetilde{\psi }\right\rangle $ represents the `spin-flip'
of $\left|\psi \right\rangle $, \emph{i.e.}, $\left|\widetilde{\psi }\right\rangle \equiv \sigma _{y}\otimes \sigma _{y}\left|\psi ^{*}\right\rangle$, and the star denotes complex conjugation in the standard basis.

The generalization of the concurrence to a mixed state $\rho $ of
two qubits is defined as the infimum of the average concurrence over all
possible pure state ensemble decompositions of $\rho $, defined as
convex combinations of pure states $S_{i}=\{p_{i},\psi _{i}\}$,
such that $\rho =\sum _{i}p_{i}|\psi _{i}\rangle \langle \psi _{i}|$.
In this way, $C\left(\rho \right)\equiv \inf _{S_{i}}\sum _{i}p_{i}C\left(\psi _{i}\right)=\inf _{S_{i}}\sum _{i}p_{i}\left|\left\langle \psi _{i}|\widetilde{\psi _{i}}\right\rangle \right|$.
 Wootters succeeded in deriving an analytic solution to this difficult
minimization procedure in terms of the eigenvalues of the nonHermitian operator $\rho \widetilde{\rho }$,
where the tilde again denotes the spin-flip
of the quantum state. The closed form solution for the concurrence
of a mixed state of two qubits is given by
$C\left(\rho \right)=\max \left\{ 0,\lambda _{1}-\lambda _{2}-\lambda _{3}-\lambda _{4}\right\}$,
 where the $\lambda _{i}$'s are ordered in decreasing order \cite{Wootters98}. Rungta,
\emph{et. al.,} extended this formalism by introducing an analytic
form for the concurrence of a bipartite system $AB$, with \emph{arbitrary}
dimensions $D_{A}$ and $D_{B}$, in an overall pure state
\cite{Rungta01}. The analytic form of this quantity, dubbed the I-concurrence,
is given by $C\left(\psi _{AB}\right)=\sqrt{2\nu _{A}\nu _{B}\left[1-tr\left(\rho _{A}^{2}\right)\right]}$,
where $\rho _{A}$ is the marginal density operator obtained by tracing
the joint pure state over subsystem $B$, and $\nu _{A}$ and $\nu _{B}$
are arbitrary scale factors. 

Coffman, \emph{et. al.}, defined the tangle $\tau _{2}$ for a system
of two qubits as the square of the concurrence \cite{Coffman00},
\emph{i.e., }

\begin{equation}
\tau _{2}\left(\rho \right)\equiv \max \left\{ 0,\lambda _{1}-\lambda _{2}-\lambda _{3}-\lambda _{4}\right\} ^{2}.\label{EqAnalyticTwoQubitTangle}\end{equation}
 Indeed, we may extend this definition directly to the result of Rungta, \emph{et. al.}, in order
to obtain an analytic form for the tangle $\tau$ of a bipartite system in
a pure state with arbitrary subsystem dimensions, \begin{equation}
\tau \left(\psi _{AB}\right)\equiv C^{2}\left(\psi _{AB}\right)=2\nu _{A}\nu _{B}\left[1-tr\left(\rho _{A}^{2}\right)\right].\label{EqPureITangle}\end{equation}
 However, when extending this definition to apply to a bipartite
\emph{mixed state} $\rho _{AB}$ with arbitrary subsystem dimensions, one must find the infimum of the average squared
pure state concurrence \cite{Osborne02}

\begin{eqnarray}
\tau \left(\rho _{AB}\right) & \equiv  & \inf _{S_{i}}\sum _{i}p_{i}C^{2}\left(\psi _{AB}^{\left(i\right)}\right)\nonumber \\
 & = & 2\nu _{A}\nu _{B}\inf _{S_{i}}\sum _{i}p_{i}\left\{ 1-tr\left[\left(\rho _{A}^{\left(i\right)}\right)^{2}\right]\right\}, \label{EqMixedITangle}
\end{eqnarray}
where we have used Eq. (\ref{EqPureITangle}) for the pure state
tangle with $\rho _{A}^{(i)}$ as the marginal state for the $i^{\mathrm{th}}$
term in the ensemble decomposition.

At this point we note that the scale factors $\nu _{A}$ and $\nu _{B}$
in Eqs. (\ref{EqPureITangle}) and (\ref{EqMixedITangle}), which may in general depend on the
dimensions of the subsystems $D_{A}$ and $D_{B}$ respectively, are
usually set to one so that agreement with the two qubit case is maintained,
and so that the addition of extra unused Hilbert space dimensions
has no effect on the value of the concurrence \cite{Rungta01}. We
will find in Sect. \ref{Sec_Ent_Sharing}, when we attempt
our own further generalization of the tangle formalism, that it is
useful to take advantage of this scale freedom. For now, however,
we adopt the usual convention both for the sake of clarity, and to
demonstrate exactly where in our proposed generalization this freedom
is required.

Using the definition given by Eq. (\ref{EqMixedITangle}), dubbed the
I-tangle in reference to the work of Rungta, \emph{et. al.}, Osborne
derived an analytic form for $\tau \left(\rho _{AB}\right)$ in the
case where the rank of $\rho _{AB}$ is no greater than two,

\begin{equation}
\tau \left(\rho _{AB}\right)=tr\left(\rho _{AB}\widetilde{\rho }_{AB}\right)+2\lambda _{min}^{\left(AB\right)}\left[1-tr\left(\rho _{AB}^{2}\right)\right],\label{EqAnalyticITangle}\end{equation}
 where $\widetilde{\rho }_{AB}$ now represents the universal inversion
\cite{Rungta01} of $\rho _{AB}$, and $\lambda _{min}^{\left(AB\right)}$
is the smallest eigenvalue of the $M$ matrix defined by Osborne \cite{Osborne02}.
The important point is that Eq. (\ref{EqAnalyticITangle}) yields
a closed form which, as we will see in Sect. \ref{SubSec_Single_Atom_Field},
is directly applicable to a specific bipartite partition of the two-atom
TCM.

\begin{figure}
{\centering \resizebox*{1.0\columnwidth}{!}{\includegraphics{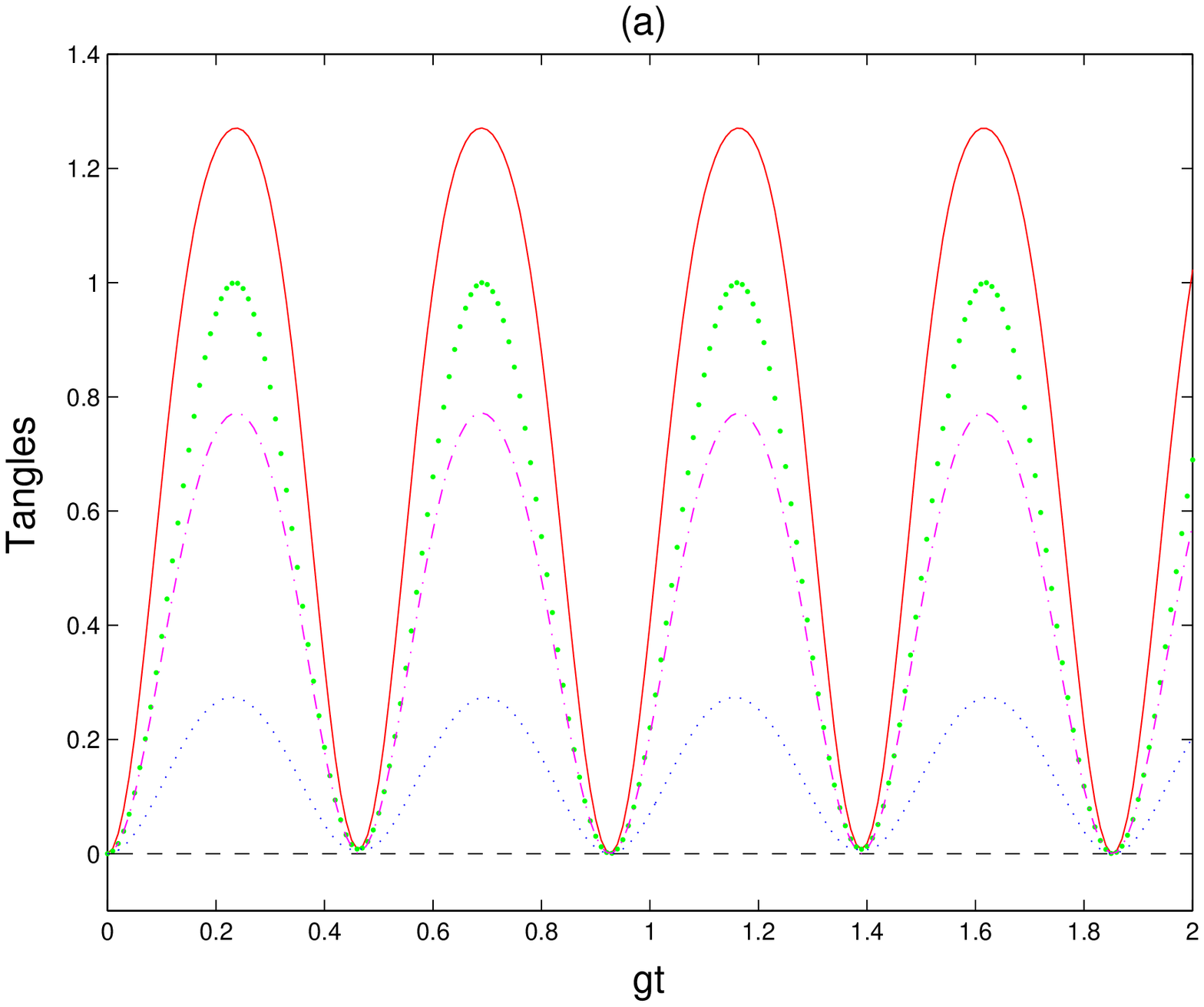}} \par}

{\centering \resizebox*{1.0\columnwidth}{!}{\includegraphics{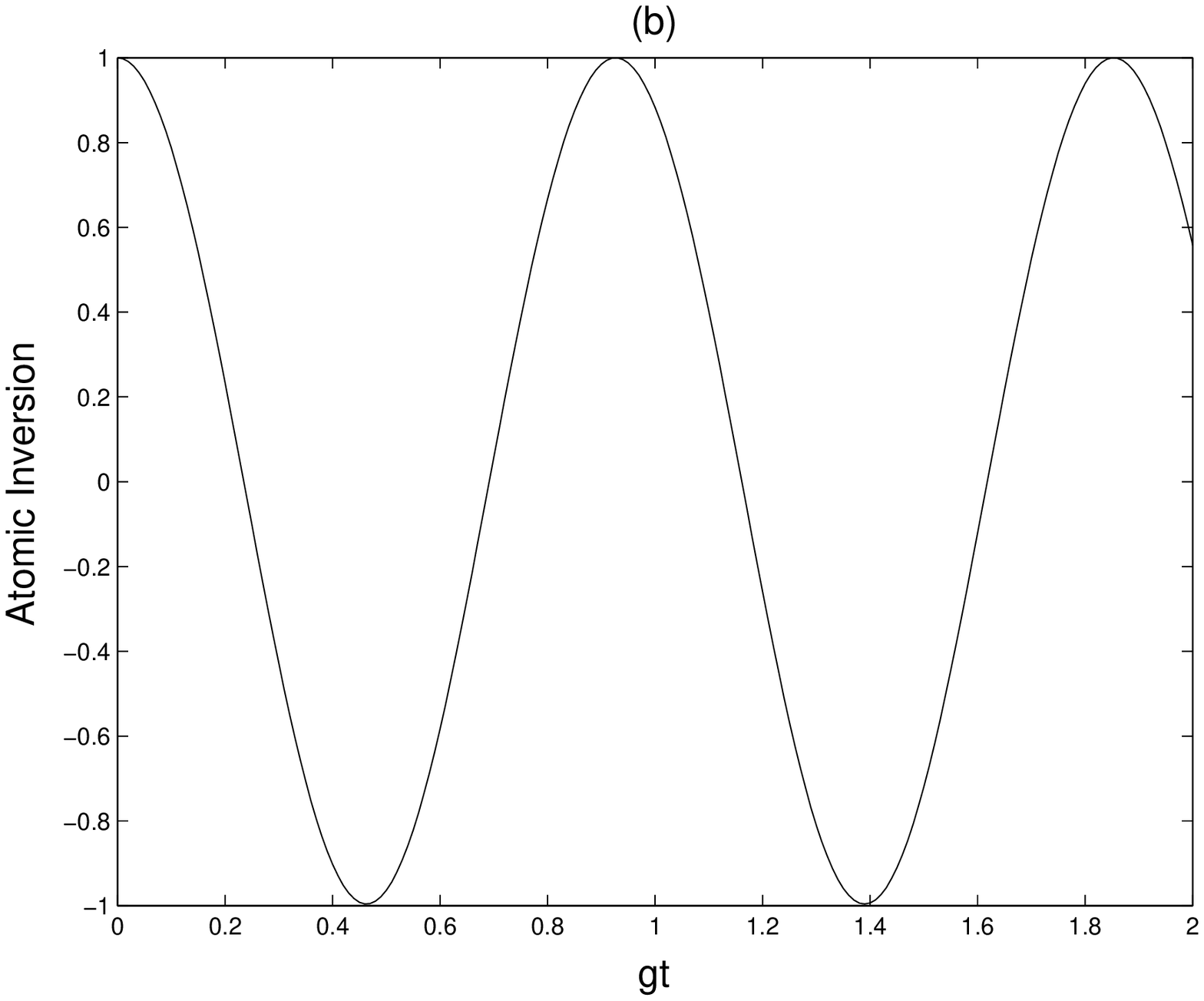}} \par}

\caption{(Color online) TCM evolution for both atoms initially in the excited state and the
field in an initial Fock state with \protect\( n=10\protect \). (a)
Solid curve (red): Field-ensemble tangle \protect\( \tau _{F\left( A_{1}A_{2}\right) }\protect \);
Large-dotted curve (green): One atom-remainder tangle \protect\( \tau _{A_{1}\left( A_{2}F\right) }\protect \);
Dashed curve (black): Atom-atom tangle \protect\( \tau _{A_{1}A_{2}}\protect \)(Note
that the atom-atom tangle is always zero for the given initial condition);
Dot-dashed curve (pink): Single atom-field tangle \protect\( \tau _{A_{1}F}\protect \);
Dotted curve (blue): Residual tangle \protect\( \tau _{A_{1}A_{2}F}\protect \).
(b) Atomic inversion of the ensemble. \label{Fig_Init_Fock}}
\end{figure}

\section{Bipartite Tangles in the Two-Atom TCM\label{Sec_Bipartite_Tangles}}

Let the two atoms in the ensemble be denoted by \( A_{1} \) and \( A_{2} \),
respectively, and the field, or quantum bus, by \( F \). Because
of the assumed exchange symmetry, there are four nonequivalent
partitions of the two-atom TCM into tensor products of bipartite subsystems: (i) the field
times the two-atom ensemble, \( F\otimes (A_{1}A_{2}) \), (ii) one
atom times the remaining atom and the field, \( A_{1}\otimes (A_{2}F)\equiv A_{2}\otimes (A_{1}F) \),
(iii) the two atoms taken separately, having traced over the field,
\( A_{1}\otimes A_{2} \), and (iv) one of the atoms times the field,
having traced over the other atom, \( A_{1}\otimes F\equiv A_{2}\otimes F \).
We calculate how the tangle for each of these partitions evolves as
a function of time under TCM Hamiltonian evolution using the formalism
reviewed in Sec. \ref{Sec_TCM}. We take the initial state to be a
pure product state of the field with the atoms. We will capture the key features
of the tangle evolution by considering three classes of initial state vectors,

\begin{subequations}
\begin{equation}
\label{Eq Init Cond Fock}
\left| e\right\rangle _{A_{1}}\otimes \left| e\right\rangle _{A_{2}}\otimes \left| n\right\rangle _{F}\equiv \left| ee,n\right\rangle ,
\end{equation}

\begin{equation}
\label{Eq Init Cond Coherent}
\left| ee,\alpha \right\rangle or\left| gg,\alpha \right\rangle ,
\end{equation}
and
\begin{equation}
\label{Eq Init Cond Symmetric}
\frac{1}{\sqrt{2}}\left( \left| eg\right\rangle +\left| ge\right\rangle \right) \otimes \left| \alpha \right\rangle or\, \frac{1}{\sqrt{2}}\left( \left| gg\right\rangle +\left| ee\right\rangle \right) \otimes \left| \alpha \right\rangle ,
\end{equation}
\end{subequations}
where \( \left| g(e)\right\rangle  \) denotes the ground (excited)
state of the atom, \( \left| n\right\rangle  \) denotes a Fock state
field with \( n \) photons, and \( \left| \alpha \right\rangle  \)
denotes a coherent state field with an average number of photons given
by \( \left\langle n\right\rangle  \). The alternatives in Eqs. (\ref{Eq Init Cond Coherent})
and (\ref{Eq Init Cond Symmetric}) arise from the fact that, in the
limit of large \( \left\langle n\right\rangle  \), the evolution
of \emph{all} of the tangles are found to be identical for the two
different initial atomic conditions, as we will see below.

\subsection{Field-Ensemble and One Atom-Remainder Tangles\label{SubSec_Field_Ensemble}}

Under the assumption that the system is in an overall pure state,
we may easily calculate the tangles in partitions (i) and (ii) above
by applying Eq. (\ref{EqPureITangle}), with \( \nu _{A}=\nu _{B}=1 \).  Specifically,

\begin{equation}
\label{Eq Field Ensemble Tangle}
\tau _{F\left( A_{1}A_{2}\right) }=2\left[ 1-tr\left( \rho _{F}^{2}\right) \right] =2\left[ 1-tr\left( \rho _{A_{1}A_{2}}^{2}\right) \right] ,
\end{equation}
 and

\begin{equation}
\label{Eq One Atom Rest Tangle}
\tau _{A_{1}\left( A_{2}F\right) }=2\left[ 1-tr\left( \rho _{A_{1}}^{2}\right) \right] =2\left[ 1-tr\left( \rho _{A_{2}F}^{2}\right) \right] ,
\end{equation}
where we have used the fact that the (nonzero) eigenvalue spectra
of the two marginal density operators for a bipartite division of
a pure state are identical \cite{Nielsen00,Ekert95} in obtaining the
rightmost equalities. These tangles have implications for the quantum
control of atomic ensembles. Because the overall system is pure, any
correlation between the field and the ensemble is necessarily in one-to-one
correspondence with the amount of entanglement between these two subsystems.
The quantum backaction on the ensemble due to measurement of the field
is thus quantified by Eq. (\ref{Eq Field Ensemble Tangle}). Alternatively,
a measurement of one atom leads to backaction on the remaining subsystem
as described by Eq. (\ref{Eq One Atom Rest Tangle}).

\begin{figure}
{\centering \resizebox*{1.0\columnwidth}{!}{\includegraphics{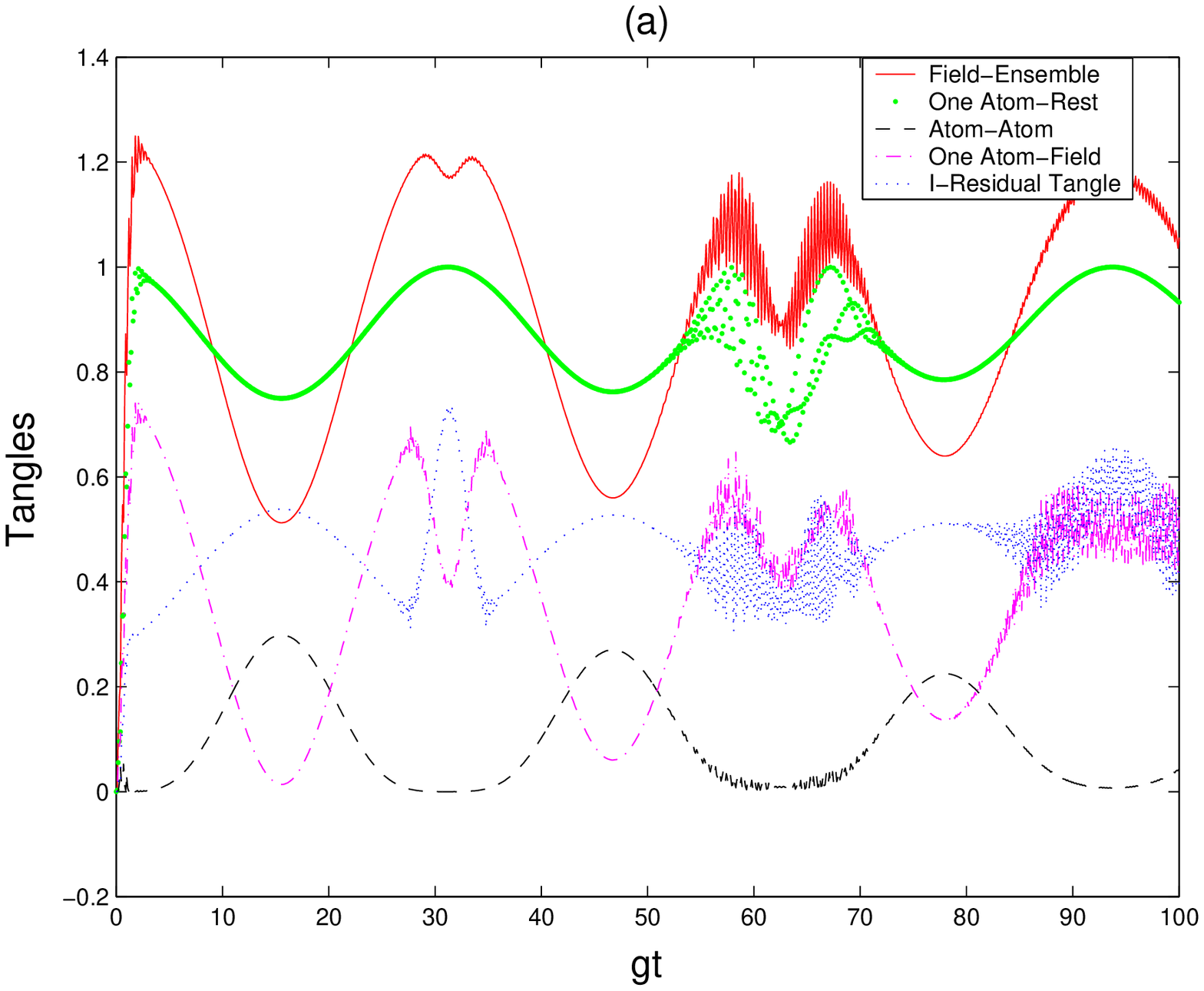}} \par}

{\centering \resizebox*{1.0\columnwidth}{!}{\includegraphics{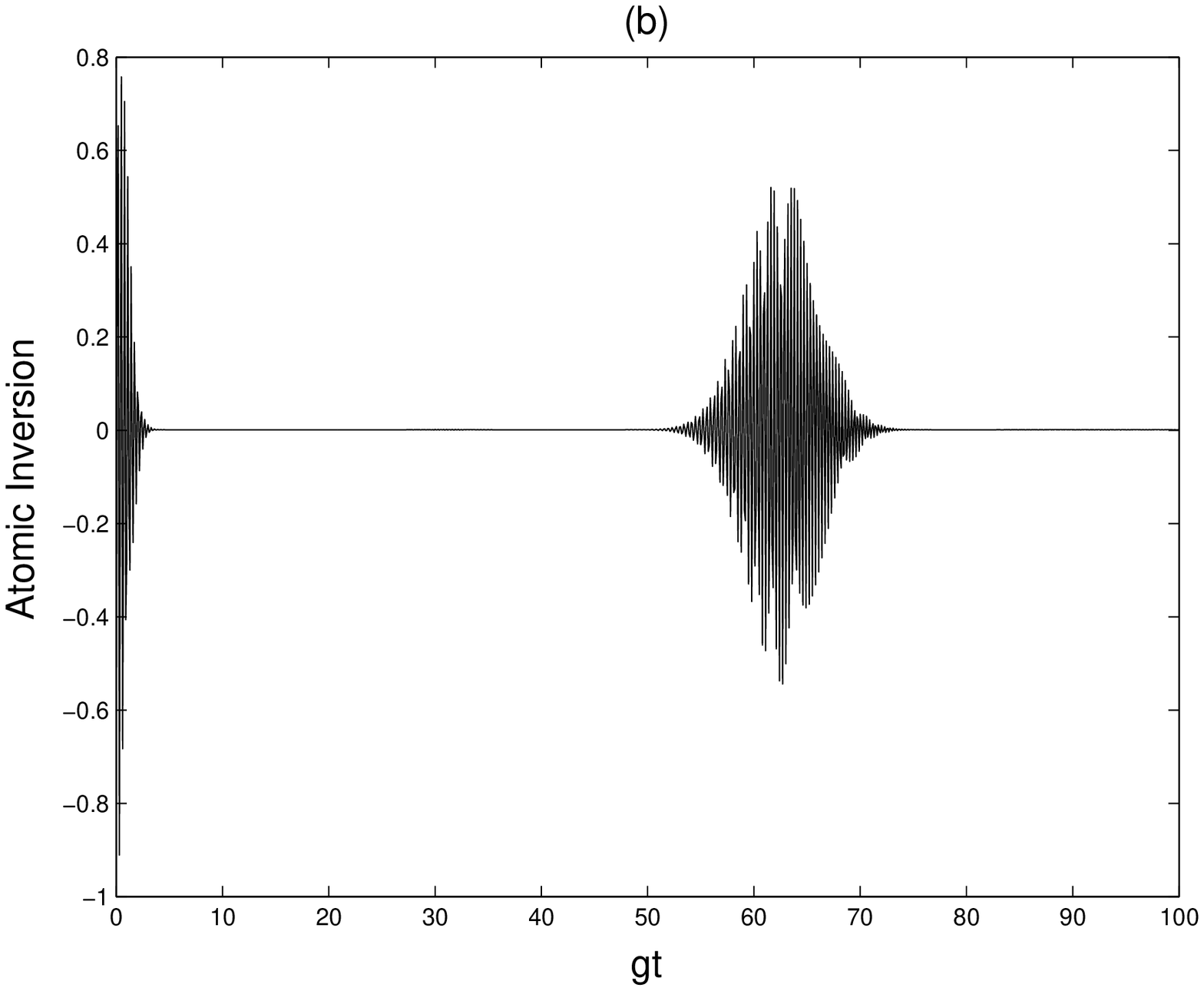}} \par}

\caption{(Color online) TCM evolution for both atoms initially in a stretched state and the
field in an initial coherent state with \protect\( \left\langle n\right\rangle =100.\protect \)
(a) Solid curve (red): Field-ensemble tangle \protect\( \tau _{F\left( A_{1}A_{2}\right) }\protect \);
Large-dotted curve (green): One atom-remainder tangle \protect\( \tau _{A_{1}\left( A_{2}F\right) }\protect \);
Dashed curve (black): Atom-atom tangle \protect\( \tau _{A_{1}A_{2}}\protect \);
Dot-dashed curve (pink): Single atom-field tangle \protect\( \tau _{A_{1}F}\protect \);
Dotted curve (blue): Residual tangle \protect\( \tau _{A_{1}A_{2}F}\protect \).
(b) Atomic inversion of the ensemble.\label{Fig_Stretched} }
\end{figure}

\begin{figure}
{\centering \resizebox*{1.0\columnwidth}{!}{\includegraphics{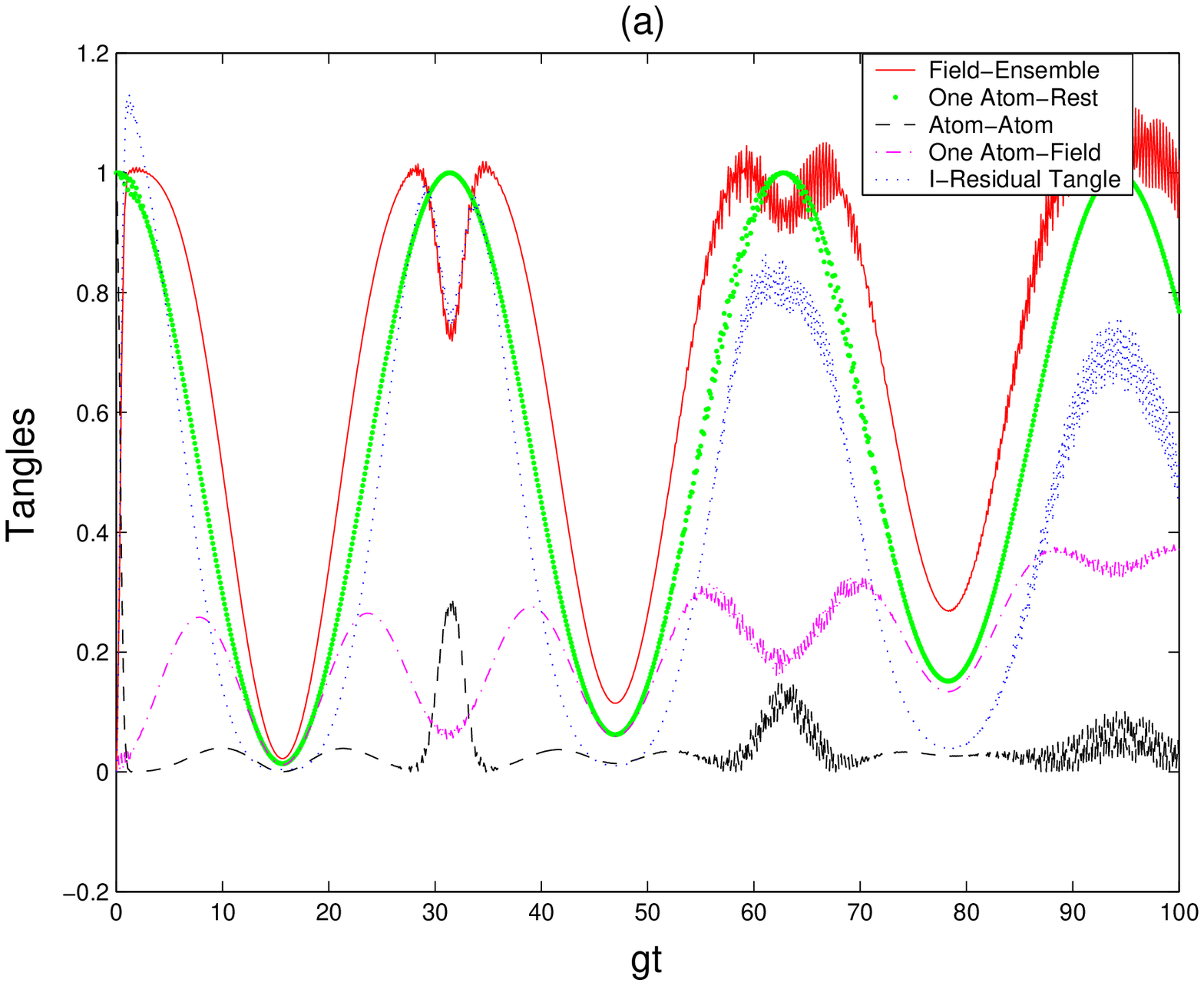}} \par}

{\centering \resizebox*{1.0\columnwidth}{!}{\includegraphics{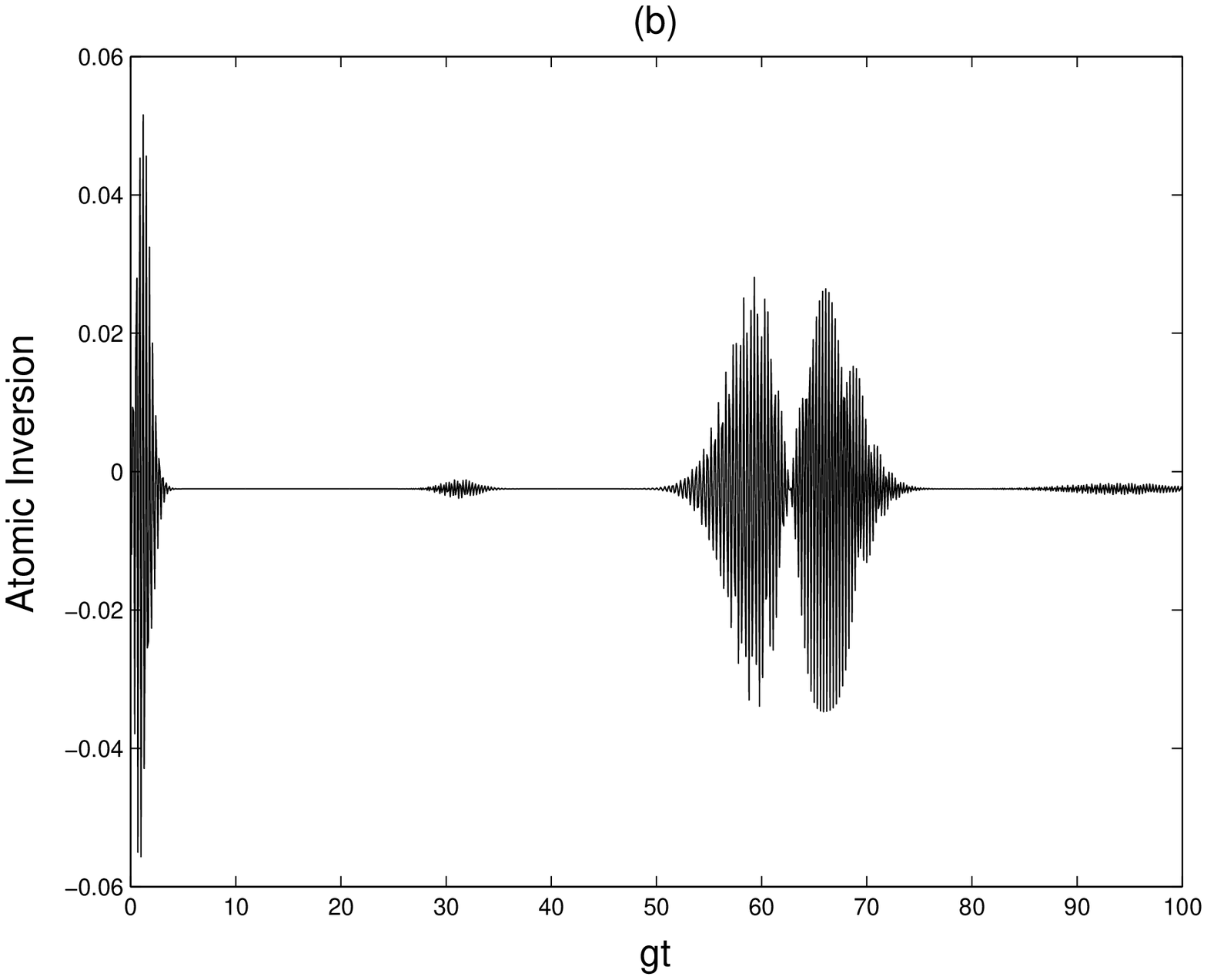}} \par}

\caption{(Color online) TCM evolution for the atoms initially in one of the symmetric states
and the field in an initial coherent state with \protect\( \left\langle n\right\rangle =100.\protect \)
(a) Solid curve (red): Field-ensemble tangle \protect\( \tau _{F\left( A_{1}A_{2}\right) }\protect \);
Large-dotted curve (green): One atom-remainder tangle \protect\( \tau _{A_{1}\left( A_{2}F\right) }\protect \);
Dashed curve (black): Atom-atom tangle \protect\( \tau _{A_{1}A_{2}}\protect \);
Dot-dashed curve (pink): Single atom-field tangle \protect\( \tau _{A_{1}F}\protect \);
Dotted curve (blue): Residual tangle \protect\( \tau _{A_{1}A_{2}F}\protect \).
(b) Atomic inversion of the ensemble.\label{Fig_Symmetric}}
\end{figure}

The time evolutions for each of the different tangles, corresponding
to the initial conditions given by Eqs. (\ref{Eq Init Cond Fock})
- (\ref{Eq Init Cond Symmetric}), are shown in Figs. \ref{Fig_Init_Fock}(a)
- \ref{Fig_Symmetric}(a) respectively. Figs. \ref{Fig_Init_Fock}(b)
- \ref{Fig_Symmetric}(b) show the time evolution of the atomic inversion,
defined as the probability of finding both atoms in the excited state
minus the probability of finding both atoms in the ground state, for
reference purposes. We find, under certain conditions, that the two
stretched states in Eq. (\ref{Eq Init Cond Coherent}) lead to identical
evolution for the tangles in \emph{all} of the bipartite partitions
of the system, corresponding to the curves shown in Fig. \ref{Fig_Stretched}(a).
Similarly, the two symmetric states given in Eq. (\ref{Eq Init Cond Symmetric})
both yield the curves shown in Fig. \ref{Fig_Symmetric}(a). This
behavior can be derived under a set of highly accurate approximations.
In the limit of large average photon number, an initial coherent state
field with zero phase will remain approximately separable from the atomic ensemble
in an eigenstate of \( J_{x}\equiv J_{+}+J_{-} \) up to times on
the order of $\left\langle n\right\rangle/{g}$ where,
in the pseudospin picture, \( J_{\pm }\equiv \sigma ^{(1)}_{\pm }+\sigma ^{(2)}_{\pm } \).
This follows immediately from the time evolution operator generated by \(H_{int}\) in Eq. (\ref{Eq TCM Hamiltonian}) in the interaction picture.
The key observation is that, for a macroscopic field, the removal or addition
of a single photon has a negligible effect. This allows one to approximate
the time evolution operator  by \(
\exp ^{-iH_{int}t/\hbar }\approx \exp ^{-ig\sqrt{\left\langle n\right\rangle }J_{x}t}\). Thus, the
eigenstates of \( J_{x} \) form a convenient basis to use in describing
the state of the atomic ensemble. This approach was taken by Gea-Banacloche
in analyzing the behavior of the single atom Jaynes-Cummings model
\cite{Gea-Banacloche90} and the generation of macroscopic superposition
states \cite{Gea-Banacloche91}, and extended to the multi-atom TCM
by Chumakov, \emph{et. al.}, \cite{Chumakov94, Delgado98, Saavedra98}.

We take as the appropriate basis the three symmetric eigenstates of
\( J_{x} \), which we label by \( m= \) -1, 0, and 1; the singlet
state, \( J=0 \), is a dark state and thus does not couple to the
field. Writing the initial state of the system as

\begin{equation}
\label{Eq Init Jx Expansion}
\left| \psi (0)\right\rangle =\sum _{m=-1}^{1}d_{m}\left| m\right\rangle \otimes \left| \alpha \right\rangle ,
\end{equation}
and using the factorization approximation \cite{Chumakov94}, we find
that the state of the system up to times on the order of $\left\langle n\right\rangle/{g}$
is given by

\begin{equation}
\label{Eq Time Evolved Factorization}
\left| \psi (t)\right\rangle \approx \sum _{m=-1}^{1}d_{m}\left| A_{m}(t)\right\rangle \otimes \left| \phi _{m}(t)\right\rangle ,
\end{equation}
 where \( \left| A_{m}(t)\right\rangle  \) and \( \left| \phi _{m}(t)\right\rangle  \)
are the time-evolved atomic and field states, respectively. The marginal
density operator for the two atoms is then

\begin{equation}
\label{Eq Two Atom Marginal Factorization}
\rho _{A_{1}A_{2}}(t)\approx \sum _{l,m}d_{l}^{*}d_{m}\left| A_{m}(t)\right\rangle \left\langle A_{l}(t)\right| f_{ml}\left( gt,\left\langle n\right\rangle \right) ,
\end{equation}
 where \( f_{ml}\left( gt,\left\langle n\right\rangle \right) \equiv \sum _{n}\left\langle n\mid \phi _{m}(t)\right\rangle \left\langle \phi _{l}(t)\mid n\right\rangle  \).
We find that this function has {}``memory\char`\"{} only for \( t\ll \sqrt{\left\langle n\right\rangle }/{g} \),
and behaves very much like a delta function for longer time scales.
Effectively, the large dimensional Hilbert space of the field acts
as a broadband reservoir for the atoms -- the generalization of the
familiar {}``collapse\char`\"{} phenomenon in the Jaynes-Cummings
model. This {}``Markoff'' approximation is valid up to times on
the order of \( 2\pi \sqrt{\left\langle n\right\rangle }/{g} \),
corresponding to the well-known revival time in the Jaynes-Cummings
Model \cite{Gea-Banacloche90}. Making this approximation in Eq. (\ref{Eq Two Atom Marginal Factorization}),
the states \( \left| A_{m}\left( t\right) \right\rangle  \) act effectively
as a {}``pointer basis\char`\"{} for decoherence \cite{Paz02} of
the atomic density matrix, \emph{i.e.},

\begin{equation}
\label{Eq Two Atom Marginal Fact and Delta}
\rho _{A_{1}A_{2}}(t)\approx \sum _{m}\left| d_{m}\right| ^{2}\left| A_{m}(t)\right\rangle \left\langle A_{m}(t)\right| .
\end{equation}
 Substituting this formula into Eq. (\ref{Eq Field Ensemble Tangle})
yields

\begin{equation}
\label{Eq Field Ens Tangle Fact and Delta}
\tau _{F\left( A_{1}A_{2}\right) }(t)\approx 2\left\{ 1-\frac{1}{4}\left[ c-h(t')\right] \right\} ,
\end{equation}
 where

\begin{eqnarray}
c & \equiv  & 4\left(\left|d_{-1}\right|^{4}+\left|d_{0}\right|^{4}+\left|d_{1}\right|^{4}\right)+2\left|d_{0}\right|^{2}\left|d_{1}\right|^{2}\nonumber \\
 &  & +\left|d_{-1}\right|^{2}\left(2\left|d_{0}\right|^{2}+3\left|d_{1}\right|^{2}\right)-4\left|d_{-1}\right|^{2}\left|d_{1}\right|^{2}\label{Eq Constant Piece}
\end{eqnarray}

\begin{eqnarray}
h\left(t'\right) & \equiv  & 2\left|d_{0}\right|^{2}\left(\left|d_{-1}\right|^{2}+\left|d_{1}\right|^{2}\right)\cos \left(4t'\right)\nonumber \\
 &  & +\left|d_{-1}\right|^{2}\left|d_{1}\right|^{2}\cos \left(8t'\right),\label{Eq Time Dependent Piece}
\end{eqnarray}
and

\begin{equation}
\label{Eq Scaled Time}
t'\equiv \frac{gt}{2\sqrt{\left\langle n\right\rangle -\frac{N}{2}+\frac{1}{2}}}.
\end{equation}

Under the factorization and Markoff approximations, the field-ensemble
tangle is given by a constant term \( c \) that depends only on the
\emph{initial} probabilities to find the atomic ensemble in each of the \( J_{x} \)
eigenstates, and a time-dependent piece \( h(t') \). These probabilities
depend solely on the absolute squares of the expansion coefficients
of the initial atomic state given by Eq. (\ref{Eq Init Jx Expansion}).
It is now clear why certain initial atomic conditions result in identical
evolution for the different tangles. For example, the initial atomic
states \( \left| gg\right\rangle  \) and \( \left| ee\right\rangle  \)
both satisfy

\begin{equation}
\label{Eq Inverted Expansion Coeff}
\left| d_{-1}\right| =\left| d_{1}\right| =\frac{1}{2}\, and\, \left| d_{0}\right| =\frac{1}{\sqrt{2}},
\end{equation}
 corresponding to identical evolution for all of the tangles shown
in Fig. \ref{Fig_Stretched}(a). Similarly, the initial atomic states
\( {1}/\sqrt{2}\left( \left| eg\right\rangle +\left| ge\right\rangle \right)  \)
and \( {1}/\sqrt{2}\left( \left| ee\right\rangle +\left| gg\right\rangle \right)  \)
both satisfy 

\begin{equation}
\label{Eq Symmetric Expansion Coeff}
\left| d_{-1}\right| =\left| d_{1}\right| =\frac{1}{\sqrt{2}}\, and\, \left| d_{0}\right| =0,
\end{equation}

corresponding to the curves shown in Fig. \ref{Fig_Symmetric}(a).
More generally, this property holds for any class of initial states
\( \left| \psi ^{(i)}\left( 0\right) \right\rangle  \) having the
form of Eq. (\ref{Eq Init Jx Expansion}) such that \( \left| d_{m}^{(i)}\right| =\left| d^{(j)}_{m}\right|  \),
\( m\in \left\{ -1,0,1\right\}  \). One
immediate consequence of this result is that the relative phase information
encoded in the initial state of the atomic system is irrelevant to
the evolution of the field-ensemble tangle.

\begin{figure}
{\centering \resizebox*{1.0\columnwidth}{!}{\includegraphics{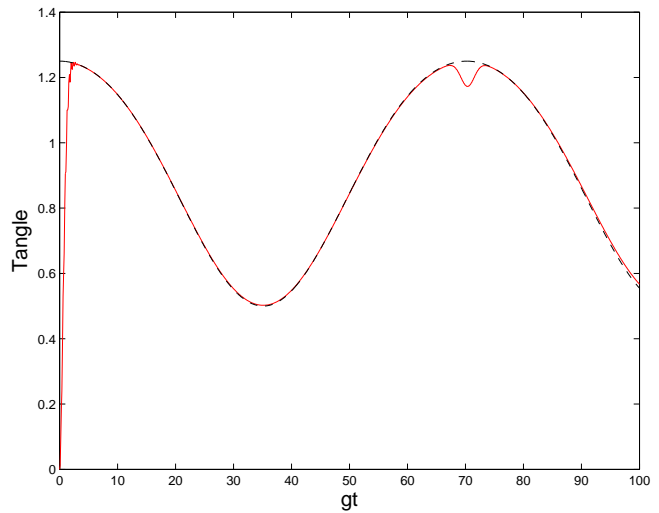}} \par}

\caption{\label{Fig_Approx}(Color online) Exact field-ensemble tangle: Solid (red) curve,
and approximate formula: Dashed (black) curve, as functions of time.}
\end{figure}

The field-ensemble tangle calculated according to Eq. (\ref{Eq Field Ensemble Tangle})
and the approximation given by Eq. (\ref{Eq Field Ens Tangle Fact and Delta})
for an initial stretched atomic state and an initial coherent state
field with \( \left\langle n\right\rangle =500 \) are shown by the
solid (red) and dashed (black) curves in Fig. \ref{Fig_Approx}, respectively.
The approximation is seen to track the exact evolution extremely well
over the range of its validity. The discrepancy at very small times
is explained by the fact that at these times the Markoff approximation
breaks down. It is also seen that our approximate solution does not
capture the small dip in the field-ensemble tangle occurring at \( t=\pi \sqrt{\left\langle n\right\rangle }/{g} \).
The absence of this feature can be explained by noting that in making
the Markoff approximation we have effectively wiped out any information
regarding the initial coherence between the \( \left| m=-1\right\rangle  \)
and \( \left| m=1\right\rangle  \) states. The presence of this dip
is then seen to be dependent upon the existence of this coherence.
This is borne out by the fact that the dip in the field-ensemble tangle
in Fig. \ref{Fig_Stretched}(a) is much shallower than that in Fig.
\ref{Fig_Symmetric}(a), where the initial atomic expansion coefficients
are given by Eqs. (\ref{Eq Inverted Expansion Coeff}) and (\ref{Eq Symmetric Expansion Coeff}),
respectively.

\subsection{Atom-Atom Tangle\label{SubSec_Atom_Atom}}

Given an initial state, we time-evolve the system according to the dynamics governed by Eq. (\ref{Eq TCM Hamiltonian}), and then trace over the field
subsystem.  The tangle of the two-atom mixed state \( \rho _{A_{1}A_{2}}\left( t\right)  \)
may then be calculated according to Eq. (\ref{EqAnalyticTwoQubitTangle}).
The resulting atom-atom tangles corresponding to the initial
conditions in Eqs. (\ref{Eq Init Cond Fock}) - (\ref{Eq Init Cond Symmetric})
are depicted by the dashed (black) curves in Figs. \ref{Fig_Init_Fock}(a)
- \ref{Fig_Symmetric}(a), respectively. These curves yield direct
insight into the state of the atomic ensemble as a function of time.  Specifically,
the atom-atom tangle quantifies the degree to which the ensemble behaves as a
collective entity, rather than as two individual particles. 

It is somewhat surprising that for the initial condition given by
Eq. (\ref{Eq Init Cond Fock}), \emph{i.e.,} when the field is
initially in a \emph{Fock state} with any value for \( n \), the
atom-atom tangle remains zero at all times, whereas the evolution
of the atom-atom tangle resulting from an initial \emph{coherent
state} field is nontrivial and, in general, nonzero. 
As a first step towards understanding these observations, we have
studied the evolution of the atom-atom tangle for other initial conditions.   When the field is initially in a Fock state and both atoms start in the ground state, the loss of an excitation in the field can result in the creation of an excitation in the atomic ensemble.  This produces entanglement between the field and the ensemble and in the single atom-field and one atom-remainder partitions.  Since it is not possible to distinguish in which atom the excitation is created, the two atoms become entangled with each other as well.  It is found that the atom-atom entanglement falls off as \( 1/{n}^2 \) so that, in the limit of a highly excited Fock state, these initial conditions yield results reminiscent of those found in Fig. \ref{Fig_Init_Fock}(a).  Specifically, we find that the entanglement in all of the different subsystem partitions always oscillate in phase at twice the Rabi frequency, and that the atom-atom tangle approaches zero as \( {n} \) becomes large.

Next, we considered the case when both atoms initially reside in a stretched state, and the initial field state consists of a coherent superposition of two neighboring Fock states. We find, on a time scale much longer than that given by the inverse of the associated Rabi frequencies, that the overall behavior again closely resembles the evolution seen for an initial field consisting of a single Fock state.  Specifically, we find that the general features of all of the different bipartite tangles oscillate in phase with one another.  However, on much shorter time-scales, the effects of dephasing between the two Rabi frequencies become apparent, yielding the first clues regarding how the observed coherent state behavior arises in terms of initial Fock state superpositions.  For example, it seems likely that these observations provide insight into the Fock-like behavior seen in Fig. \ref{Fig_Stretched}(a) at the revival time, when there is a partial rephasing of the Rabi frequencies. At this time all of the bipartite tangles decrease simultaneously, while at other times the tangles in certain bipartite partitions may be completely out of phase with one another.  We are currently performing a detailed study of the entanglement that can be dynamically generated between the two atoms under TCM evolution for the most general initial conditions in order to better understand this behavior.

\subsection{Single Atom-Field Tangle\label{SubSec_Single_Atom_Field}}

The final bipartite partition of the two-atom TCM is that consisting
of a single atom, say \( A_{1} \), as one subsystem and the field
\( F \) as the second subsystem. Again by exchange symmetry \( \tau _{A_{1}F}=\tau _{A_{2}F} \),
so we need calculate only one of these quantities. Because the tripartite
system is in an overall pure state, the Schmidt decomposition theorem
\cite{Nielsen00,Ekert95} implies that the marginal density operator \( \rho _{A_{1}F} \)
has at most rank two. The rank of the reduced density matrix is set
by the dimension of the smallest subsystem, which in this case is a two-level atom. This is exactly the scenario envisioned by
Osborne \cite{Osborne02}, as described in Sec. \ref{Sec_Tangle_Formalism}.
The tangle corresponding to this partition, \( A_{1}\otimes F \),
is computed by first tracing over the state of the remaining atom,
\( A_{2} \), and then applying Eq. (\ref{EqAnalyticITangle}).

Employing this procedure,

\begin{equation}
\label{Eq One Atom Field Tangle}
\tau _{A_{1}F}=tr\left( \rho _{A_{1}F}\widetilde{\rho }_{A_{1}F}\right) +2\lambda ^{\left( A_{1}F\right) }_{min}\left[ 1-tr\left( \rho _{A_{1}F}^{2}\right) \right] ,
\end{equation}
where \( \lambda _{min}^{\left( A_{1}F\right) } \) represents the
minimum eigenvalue of the Osborne \( M \) matrix \cite{Osborne02}
generated from the marginal density operator \( \rho _{A_{1}F} \).
The dot-dashed (pink) curves in Figs. \ref{Fig_Init_Fock}(a)
- \ref{Fig_Symmetric}(a) give the time evolution of the single atom-field
tangle for the different initial conditions considered.

We are now in possession of closed forms for the tangles of all
bipartite partitions of the two-atom TCM. Any other entanglement that
the system may possess must necessarily be in the form of irreducible
three-body quantum correlations. In section \ref{Sec_Ent_Sharing}
we review the \emph{residual tangle} formalism introduced by Coffman, \emph{et.
al.}, in order to quantify this type of tripartite entanglement in
a system of three qubits. We then propose a generalization of this
quantity that is applicable to a \( 2\otimes 2\otimes D \) system
in an overall pure state. This extension of the tangle formalism allows us to study
the phenomenon of entanglement sharing in the two-atom TCM.

\section{Entanglement Sharing and the Residual Tangle\label{Sec_Ent_Sharing}}

The concept of entanglement sharing studied in \cite{Coffman00, Dennison01}
refers to the fact that entanglement cannot be freely distributed
among subsystems in a multipartite, \emph{i.e.}, tripartite or higher,
system. Rather, the distribution of entanglement in these systems
is subject to certain constraints. As a simple example, consider a
tripartite system of three qubits labeled \( A, \) \( B, \) and
\( C. \) Suppose that qubits \( A \) and \( B \) are known to be
in a maximally entangled pure state, \emph{e.g.,} the singlet state,
given by $\left|\psi _{AB}\right\rangle =1/\sqrt{2}\left(\left|01\right\rangle -\left|10\right\rangle \right)$ when written in the logical basis. In this case, it is obvious that
the overall system \( ABC \) is constrained such that no entanglement
may exist either between \( A \) and \( C \) or between \( B \)
and \( C. \) Otherwise, tracing over subsystem \( C \) would necessarily
result in a \emph{mixed} marginal density operator for \( AB \) in
contradiction to the known purity of the singlet state.

Coffman, \emph{et. al.}, analyze the phenomenon of entanglement sharing
for a system of three qubits in an overall pure state in full generality
by introducing a quantity known as the residual tangle \cite{Coffman00}.
This definition is motivated by the observation that the tangle of
\( A \) with \( B \) plus the tangle of \( A \) with \( C \) cannot
exceed the tangle of \( A \) with the joint subsystem \( BC \),
\emph{i.e.},

\begin{equation}
\label{Eq Tangle Ineq}
\tau _{AB}+\tau _{AC}\leq \tau _{A\left( BC\right) }.
\end{equation}
 Here, \( \tau _{AB} \) and \( \tau _{AC} \) are calculated according
to Eq. (\ref{EqAnalyticTwoQubitTangle}), and \( \tau _{A\left( BC\right) } \)
may be obtained from Eq. (\ref{EqPureITangle}).

The original proof \cite{Coffman00} of the inequality in Eq. (\ref{Eq Tangle Ineq}),
which forms the heart of the phenomenon of entanglement sharing for
the case of three qubits, may be substantially simplified by making
use of certain results due to Rungta, \emph{et. al.} Specifically,
we note that \cite{Rungta01}

\begin{equation}
\label{Eq Trace Rho Rho Tilde}
tr\left( \rho _{xy}\widetilde{\rho }_{xy}\right) =1-tr\left( \rho _{x}^{2}\right) -tr\left( \rho _{y}^{2}\right) +tr\left( \rho _{xy}^{2}\right) \geq 0
\end{equation}
 for subsystems \( x \) and \( y \) having arbitrary Hilbert space dimensions.
Under the assumption that \( x \) and \( y \) are in an overall
pure state with a third subsystem \( z \), Eq. (\ref{Eq Trace Rho Rho Tilde})
may be rewritten

\begin{equation}
\label{Eq Trace Rho Rho Tilde Pure}
tr\left( \rho _{xy}\widetilde{\rho }_{xy}\right) =1-tr\left( \rho _{x}^{2}\right) -tr\left( \rho _{y}^{2}\right) +tr\left( \rho _{z}^{2}\right) \geq 0,
\end{equation}
 where we have used the equality of the nonzero eigenvalue spectra
of \( \rho _{xy} \) and \( \rho _{z} \). Then, by the observation
\cite{Coffman00} that for an arbitrary state of two qubits \( A \)
and \( B, \) the following upper bound on the tangle defined by Eq.
(\ref{EqAnalyticTwoQubitTangle}) holds\begin{equation}
\label{Eq Tangle Upper Bound}
\tau _{2}\left( \rho _{AB}\right) \leq tr\left( \rho _{AB}\widetilde{\rho }_{AB}\right) ,
\end{equation}
 and by Eq. (\ref{EqPureITangle}) with \( \nu _{A}=\nu _{B}=1 \),
the inequality in Eq. (\ref{Eq Tangle Ineq}) follows immediately.

Subtracting the terms on the left hand side of Eq. (\ref{Eq Tangle Ineq})
from that on the right hand side yields a positive quantity referred
to as the \emph{residual tangle \( \tau _{ABC} \), i.e.},

\begin{equation}
\label{Eq Residual Tangle}
\tau _{ABC}\equiv \tau _{A\left( BC\right) }-\tau _{AB}-\tau _{AC}.
\end{equation}
 The residual tangle is interpreted as quantifying the inherent \emph{tripartite}
entanglement present in a system of three qubits, \emph{i.e.,} the
entanglement that cannot be accounted for in terms of the various
bipartite tangles. This interpretation is given further support by
the observation that the residual tangle is invariant under all possible
permutations of the subsystem labels \cite{Coffman00}.

We wish to generalize the residual tangle, defined for a system of
three qubits, to apply to a \( 2\otimes 2\otimes D \) quantum system
in an overall pure state so that we may study entanglement sharing
in the two-atom TCM. Note that we already have all of the other tools
needed for such an analysis. Specifically, from section \ref{Sec_Bipartite_Tangles},
we know the analytic forms for all of the different possible bipartite
tangles in such a system.

Any proper generalization of the residual tangle must, at a minimum,
be a positive quantity, and be equal to zero if and only if there
is no tripartite entanglement in the system, \emph{i.e.,} if and only
if all of the quantum correlations can be accounted for using only
bipartite tangles. It should also reduce to the definition of the
residual tangle in the case of three qubits. Further it is reasonable
to require, if this is to be a true measure of irreducible three-body
correlations, that symmetry under permutation of the subsystems
be preserved, and that it remain invariant under local unitary operations.  Finally, 
we conjecture that this quantity satisfies the requirements
for being an entanglement monotone \cite{Vedral97,Vidal00} under the set of stochastic
local operations and classical communication (SLOCCs), or equivalently, under the set of invertible local operations \cite{Dur00}.  We limit the monotonicity requirement to this restricted set of operations since, in the context of entanglement sharing, we are only concerned with LOCCs that preserve the local ranks of the marginal density operators such that all subsystem dimensions remain constant.

Let \( A \) and \( B \) again be qubits, and let \( C \) now be
a \( D \)-dimensional system with the composite system \( ABC \)
in an overall pure state. We note that, under these assumptions, we
are still capable of evaluating each of the terms on the right hand
side of Eq. (\ref{Eq Residual Tangle}) analytically using the results of section \ref{Sec_Bipartite_Tangles}.  However,
we cannot simply use the definition of the residual tangle (with \( C \)
now understood to represent a \( D \)-dimensional system) as the
proper generalization for two reasons. First, since the three subsystems
are no longer of equal dimension, symmetry under permutations of the
subsystems is lost. However, as we will see, this problem is easily
fixed by explicitly enforcing the desired symmetry. The second, and
more difficult problem to overcome is the fact that the inequality
given by Eq. (\ref{Eq Tangle Ineq}) no longer holds for our generalized
system because \( \lambda _{min} \) in Eq. (\ref{EqAnalyticITangle})
can be negative, implying that Eq. (\ref{Eq Residual Tangle}) can
also be negative.

The required permutation symmetry may be restored by taking our generalization
of the residual tangle, which we dub the \emph{I-residual tangle}
in reference to previous work and continue to denote by \( \tau _{ABC} \),
to be

\begin{eqnarray}
\tau _{ABC} & \equiv  & \frac{1}{3}\left(\tau _{A\left(BC\right)}+\tau _{B\left(AC\right)}+\tau _{C\left(AB\right)}\right)\nonumber \\
 &  & -\frac{2}{3}\left(\tau _{AB}+\tau _{AC}+\tau _{BC}\right)\label{Eq Residual Tangle Gen}
\end{eqnarray}
The definition in Eq. (\ref{Eq Residual Tangle Gen}) is obtained
by averaging over all possible relabelings of the subsystems in Eq.
(\ref{Eq Residual Tangle}). By inspection, it is obvious that Eq.
(\ref{Eq Residual Tangle Gen}) preserves permutation symmetry. However,
it still suffers from the problem that its value can be negative.
In order to deal with this difficulty, we make use of the arbitrary
scale factors appearing in Eqs. (\ref{EqPureITangle}) and (\ref{EqMixedITangle}),
and discussed in section \ref{Sec_Tangle_Formalism}.

Let \( d \) be the smaller of the two `dimensions' of two arbitrary
dimensional subsystems \( x \) and \( y \), \emph{i.e.}, \( d\equiv \min \left\{ D_{x},D_{y}\right\}  \).
Note that by dimension we do not necessarily mean the total Hilbert
space dimension of the physical system under consideration, but only
the number of different Hilbert space dimensions \emph{that contribute
to the formation} of the overall pure state of the system. This is
a subtle but important point which automatically enforces insights
like those due to Rungta, \emph{et. al.}, \cite{Rungta01} and Verstraete, \emph{et. al.}, \cite{Verstraete01} which state that the scale chosen for a measure of entanglement must be invariant
under the addition of extra, but unused, Hilbert space dimensions.
The two-atom TCM provides one example of the relevant physics underlying
these ideas.

Consider, for example, the bipartite partitioning of the TCM into
a field subsystem with \( D_{F}=\infty  \), and an ensemble subsystem
consisting of the two qubits with \( D_{A_{1}A_{2}}=4 \). Any entangled
state of the overall system has a Schmidt decomposition with at most
four terms, implying that the field effectively behaves like a four-dimensional
system. Further, since the Tavis-Cummings Hamiltonian given by Eq.
(\ref{Eq TCM Hamiltonian}) does not induce couplings between the
field and the singlet state of the atomic ensemble, \emph{i.e.}, the
singlet state is a dark state, the field behaves effectively as a
three-level system, or \emph{qutrit}, in the context of the TCM. Accordingly,
in any entangled state of the field with the ensemble, the field is considered
to have a dimension no greater than three. We employ this revised
definition of dimension throughout the remainder of the paper.

We now make the choice 
\begin{equation}
\label{Eq Scaling}
\nu _{A}\nu _{B}={d}/{2},
\end{equation}
when calculating
each of the bipartite tangles appearing on the right hand side of
Eq. (\ref{Eq Residual Tangle Gen}). This choice is made for several
reasons. First of all, it is in complete agreement with the two qubit
case, yielding \( \nu _{A}\nu _{B}=1 \) as required. Indeed, when
\( A \), \( B \), and \( C \) are all qubits, the residual tangle
given by Eq. (\ref{Eq Residual Tangle}) is recovered. Secondly, it
takes differences in the Hilbert space dimensions of the subsystems into
account when setting the relevant scale for each tangle. This is important
since, in order to study the phenomenon of entanglement sharing, the
tangles for each of the different bipartite partitions must be compared
on a common scale. It is reasonable that this scale be a function
of the smaller of the two subsystem dimensions since, for an overall
pure state, it is this quantity that limits the number of terms in
the Schmidt decomposition. Finally, it is conjectured that a proper
rescaling of the various tangles will result in the positivity
of Eq. (\ref{Eq Residual Tangle Gen}).

Note that when applying the proposed rescaling to the terms on the
right hand side of Eq. (\ref{Eq Residual Tangle Gen}), the only term
affected is \( \tau _{C\left( AB\right) } \), which is rescaled by
one-half of the smaller of the two subsystem dimensions \( D_{C} \)
and \( D_{AB} \). Each of the other terms remains unaltered since,
in each case, at least one of the two subsystems involved is a qubit.
The net effect of this rescaling is to increase the `weight' of the
tangle between \( C \) and \( AB \) relative to that of the rest of the
tangles. This is reasonable when one recognizes that both \( AB \),
a system of two qubits, and \( C \), a \( D \)-dimensional system
(in the case \( D>2 \)), have \emph{entanglement capacities} \cite{Dennison01}
exceeding that of a single qubit.

The requirement that the I-residual tangle be invariant under local
unitary operations follows trivially, since each term on the right
hand side of Eq. (\ref{Eq Residual Tangle Gen}) is known to satisfy
this property individually.  It is still an open question
as to whether or not the proposed rescaling is sufficient to preserve
positivity when generalizing the residual tangle, Eq. (\ref{Eq Residual Tangle}),
to the I-residual tangle, Eq. (\ref{Eq Residual Tangle Gen}).  However, numerical
calculations give strong evidence that this is the case. The I-residual
tangle has been calculated for over two-hundred million randomly generated
pure states of a \( 2\otimes 2\otimes 3 \) system and of a \( 2\otimes 2\otimes 4 \)
system, the only nontrivial possibilities. In each instance the resulting
quantity has been positive.  We conjecture that the I-residual tangle satisfies the requirements of positivity and monotonicity under SLOCC not only for a \( 2\otimes 2\otimes D \) system, where closed forms currently exist for all of the terms on the right hand side of Eq. (\ref{Eq Residual Tangle Gen}), but for the most general \( {D}_{A}\otimes {D}_{B}\otimes {D}_{C} \) dimensional tripartite system in an overall pure state (with the proper scaling of each term again given by Eq. (\ref{Eq Scaling})).  The I-residual tangle arising in the context of the two-atom 
TCM is shown by the blue curves in Figs. \ref{Fig_Init_Fock}(a) - \ref{Fig_Symmetric}(a).

The residual tangle, as well as our proposed generalization of this
quantity, may be interpreted as the irreducible tripartite entanglement
in a system since it cannot be accounted for in terms of any combination
of bipartite entanglement measures \cite{Coffman00}. A slightly different,
and possibly more enlightening interpretation is that the I-residual
tangle quantifies the amount of `freedom' that a system has in satisfying
the constraints imposed by the phenomenon of entanglement sharing.
If the I-residual tangle of a tripartite system is zero, then each bipartite
tangle is uniquely determined by the values of all of the other bipartite
tangles. Alternatively, if \( \tau _{ABC} \) is strictly greater
than zero, then the bipartite tangles enjoy a certain latitude in
the values that each may assume while still satisfying the positivity criterion. 
The larger the value of the I-residual
tangle, the more freedom the system has in satisfying
the entanglement sharing constraints. This reasoning highlights the
relationship between entanglement sharing and the I-residual tangle.

Finally, we may interpret the I-residual tangle as the \emph{average
fragility} of a tripartite state under the loss of a single subsystem.
That is, if one of the three subsystems is selected at random and
discarded (or traced over), then the I-residual tangle quantifies
the amount of \emph{bipartite} entanglement that is lost, on average.
It is the existence of physically meaningful interpretations
such as these which prompt us to postulate this new measure of tripartite
entanglement for a \( 2\otimes 2\otimes D \) system in an overall
pure state, rather than to rely on previously defined measures based
on normal forms \cite{Verstraete01} or on the method of hyperdeterminants \cite{Miyake03},
for example. At this point it is unclear what, if any connection these
entanglement monotones have to the entanglement that exists in different
bipartite partitions of the system, a key ingredient in any discussion
of entanglement sharing.

The constraint imposed by entanglement sharing on the values of the
various bipartite tangles, each of which is known to be a positive
function, is simply that Eq. (\ref{Eq Residual Tangle Gen}) cannot
be negative. It then follows that the strongest constraint of this
form is placed on the two-atom TCM when the I-residual tangle is equal
to zero. This occurs (to a good approximation) periodically in Fig.
\ref{Fig_Init_Fock}(a) for the initial condition given by
Eq. (\ref{Eq Init Cond Fock}). It is at these points that each bipartite
tangle is uniquely determined in terms of the values of all of the
other bipartite tangles. Conversely, at one half of this period when
the I-residual tangle achieves its maximum value, the various bipartite
partitions enjoy their greatest freedom with respect to how entanglement
may be distributed throughout the system while still satisfying the
entanglement sharing constraints. The distribution of correlations
is, of course, still determined by the initial state of the system
and by the TCM time evolution, both of which we consider to be separate constraints.

Similarly, the dotted (blue) curves in Figs. \ref{Fig_Stretched}(a)
and \ref{Fig_Symmetric}(a) show the evolution of the residual tangle
for the initial states given by Eqs. (\ref{Eq Init Cond Coherent})
and (\ref{Eq Init Cond Symmetric}), respectively. Note how the more
complicated behavior resulting from an initial coherent state field
arises from a specific superposition of Fock states, the tangles of
which all have a simple oscillatory evolution. This suggests that
the phenomenon of entanglement sharing may offer a useful perspective
from which to investigate the way in which the coherent state evolution
results from a superposition of Fock state evolutions.

The fact that the TCM Hamiltonian leads to a nonzero I-residual
tangle is interesting in its own right. Inspection of Eq. (\ref{Eq
TCM Hamiltonian}) shows that this model does not include a
physical mechanism, \emph{e.g.}, a dipole-dipole coupling term
enabling direct interaction between the two atoms in
the ensemble, but only for coupling between the field and the
atoms. Consequently, all interactions between the atoms are
mediated by the electromagnetic field via the exchange of photons,
and are in some sense indirect. This, however, turns out to be
sufficient to allow genuine tripartite correlations to develop in
the system as evidenced by values of the I-residual tangle that
are strictly greater than zero.

Finally, we note that we have considered an alternative approach to understanding the
constraints on the distribution of entanglement among the different subsystem
partitions of the two-atom TCM by using the relative entropy of
entanglement \cite{Vedral97} as our entanglement measure. This quantity generalizes 
in a straightforward manner to
the multipartite case \cite{vedral02}, and has a clear physical interpretation relating the
amount of entanglement in a state to its distance from the set of separable states \cite{Vedral97}. The existence of upper and lower bounds in
the tripartite case \cite{plenio98} yields another method by which to investigate the genuine
three body entanglement arising in the two-atom TCM.  The results of the relative entropy of entanglement approach will be presented in a subsequent article.

\section{Summary and Future Directions\label{Sec_Summary}}

The two-atom Tavis-Cummings model provides the simplest example of
a collection of two-level atoms, or qubits, sharing a common coupling
to the electromagnetic field. A detailed understanding of the evolution
of entanglement in different bipartite partitions of this model is
valuable for both fundamental theoretical investigations, and for
accurately describing the behavior of certain nontrivial, yet experimentally
realizable systems. Our proposed generalization of the residual tangle
augments the current formalism, and allows one to analyze the
irreducible three-body correlations that arise in a broader class of tripartite systems,
providing a tool useful for studying the phenomenon of entanglement sharing in
the context of a physically relevant and accessible system.

In future work we hope to generalize this analysis to include ensembles with an arbitrary number of atoms.  This will require further extensions of the tangle formalism in order to quantify both the entanglement in a mixed state of a bipartite system of arbitrary dimensions having a local rank greater than two, and the multipartite entanglement
in a system composed of more than three subsystems.  Ultimately, we hope to connect this analysis to the phenomenon of quantum backaction on individual particles when the whole ensemble is measured.  The tradeoff between the information gained about a system and the disturbance caused to that system is certainly fundamental to quantum mechanics \cite{Fuchs01,Banaszek01,D'Ariano02}.  However, the relationship of this tradeoff to multiparticle entanglement is far from clear.  Such an understanding would not only be a crucial step in designing protocols for the quantum control of ensembles, but would also provide deeper insight into the nature of the correlations at the heart of quantum mechanics \cite{Mermin98}.

\section{Acknowledgements} T. E. T. and I. H. D. acknowledge support from the Office of Naval Research under Contract No.~N00014-00-1-0575.  T. E. T. would also like to thank Tobias Osborne for helpful correspondence.  A. D. and I. F-G would like to thank Carlton M. Caves, Ivan H. Deutsch and the
rest of the Information Physics group at the University of New
Mexico for their hospitality. A. D. acknowledges support from
Fondecyt under Grant No. 1030671. I. F.-G. would like to thank
Consejo Nacional de Ciencia y Tecnologia (Mexico) Grant no.
115569/135963 for financial support and A. Carollo for useful
discussions.

\bibliography{MyLib}   %>>>> bibliography data in MyLib.bib

\end{document}